\documentstyle[prb,aps,preprint]{revtex}
\begin{document}
\draft
\preprint{}

\title{\bf GAPLESS PHASES IN AN S=1/2 QUANTUM SPIN CHAIN\\ WITH
        BOND ALTERNATION }
\author {A. A. Nersesyan\cite{email}}
\address{
Institute of Theoretical Physics,
Chalmers University of Technology, S-41296 G\"{o}teborg, Sweden
\\
and Institute of Physics,
Georgian Academy of Sciences, Tamarashvili 6, 380077, Tbilisi, Georgia}

\author{A.Luther}
\address
{NORDITA, Blegdamsvej 17, DK-2100 Copenhagen 0, Denmark}

\date{\today}
\maketitle
\begin{abstract}
The $S=1/2$ XXZ spin chain with the staggered XY anisotropy
$$
H =
J \sum_{n}^{N} (S^{x}_{n} S^{x}_{n+1} + S^{y}_{n} S^{y}_{n+1} +
\Delta S^{z}_{n} S^{z}_{n+1})
- \delta \sum_{n}^{N} (-1)^{n}
(S^{x}_{n} S^{x}_{n+1} - S^{y}_{n} S^{y}_{n+1})
$$
is shown to possess gapless,
Luttinger-liquid-like phases in a wide range of its parameters: the XY-like
phase
and spin nematic phases, the latter characterized by a two-spin order parameter
breaking translational and spin rotation symmetries. In the simplest,
exactly solvable case $\Delta = 0$, the spectrum remains gapless at arbitrary
$J$ and $\delta$ and is described by two massless Majorana (real) fermions with
different
velocities $v_{\pm} = |J \pm \delta|$.  At $|\delta| < J$ the staggered
XY anisotropy does not influence the ground state of the system (XY phase).
At $|\delta| > J$, due to level crossing, a spin nematic state is realized,
with
$\uparrow \uparrow \downarrow \downarrow$
and $\uparrow \downarrow \downarrow \uparrow$ local symmetry of the
$xx$ and $yy$ spin correlations. The spin correlation functions are calculated
and
the effect of thermally induced spin nematic ordering
in the XY phase ("order from disorder") is discussed.

The role of a finite $\Delta$ is studied in the limiting cases $|\delta| \ll J$
and
$|\delta| \gg J$, using bosonization method. On the basis of the derived
 self-dual field-theoretical model,
similar to the one recently proposed to treat two weakly coupled Luttinger
chains,
a confinement regime is described, in which fermionic excitations fail to be
observable
due to Luttinger-liquid effects. Kosterlitz-Thouless transitions
to massive phases,
driven by the $\delta-$anisotropy, are discussed.

\end{abstract}

\narrowtext
\section{Introduction}
There has been considerable recent interest in the
$S = 1/2$ quantum spin chains with
alternating ferromagnetic and antiferromagnetic bonds \cite{Hida,Kom-Tas,Yam}:
\begin{equation}
H = \sum_{j} [({\bf S}_{2j}\cdot {\bf S}_{2j+1})
- \beta ({\bf S}_{2j-1} \cdot {\bf S}_{2j})]. \label {eq:alt}
\end{equation}
The model (\ref{eq:alt}) continuously connects a gapful phase of
decoupled singlets
($\beta = 0$) with the
limit of the  $S = 1$ chain ($\beta \rightarrow \infty$) and
so far was mostly considered in the context of the
Haldane gap problem \cite{Hald-gap}.

In this paper, we would like to draw attention to another,
in a sense, opposite
aspect of the problem of bond-alternating spin chains.
In the rotationally (SU(2)) invariant $S=1/2$ chain
the formation of a massive phase appears as a result of
bond alternation $(\beta \neq -1)$.
On the other hand, as follows from the (Bethe ansatz) exact solution
of the translationally invariant XYZ model \cite {XYZ}, an
Ising-like mass gap is always present in the excitation
spectrum as long as continuous spin rotation symmetry remains
fully broken. One then may ask:
Are the requirements of translational and spin rotation symmetries robust
for realization of a gapless, Luttinger-liquid low-energy behavior of $S=1/2$
spin chains?
Can the bond alternation
{\it and} full breakdown of spin rotational symmetry, each
separately
generating a gap, result in a gapless (critical) regime
when acting together?
As shown below (Sec.2), among several unitarily equivalent
possibilities, the
XXZ spin chain with the staggered XY anisotropy
$$ H = H_{XXZ} + H_{\delta} $$
\begin{equation}
= J \sum_{n}^{N} (S^{x}_{n} S^{x}_{n+1} + S^{y}_{n} S^{y}_{n+1} +
\Delta S^{z}_{n} S^{z}_{n+1})
- \delta \sum_{n}^{N} (-1)^{n}
(S^{x}_{n} S^{x}_{n+1} - S^{y}_{n} S^{y}_{n+1}).
\label {eq:basic}
\end{equation}
is a nontrivial example of such a system exhibiting  gapless
phases in a wide range of its parameters with rather
interesting properties.

In Sec.3 we first consider the simplest, exactly
solvable case of the XY bond-alternating
chain ($\Delta = 0$). The Jordan-Wigner transformation reduces it
to a 1d system of spinless fermions with a half-filled band and
a Cooper pairing with the total momentum equal $\pi$.
This model is also equivalent to two decoupled
quantum Ising chains (i.e. 1d Ising models in transverse magnetic
fields) {\it at criticality}. Therefore,
for any ratio $\delta/J$, the excitation
spectrum is gapless, and elementary excitations are represented by
 two Majorana (real) fermions propagating with different
velocities $v_{\pm} = |J \pm \delta| a$.
At $|\delta| < J$ the staggered
XY anisotropy does not influence the ground state of the system
but reveals at the level of excitations, leading to splitting of
the two gapless branches of the spectrum. This affects
the time dependence of the spin correlation functions and
thermodynamics.

At $|\delta| > J$, due to
level crossing, the XY ground state is changed by
another gapless phase - a
spin nematic (SN) state, characterized by
broken translational and spin rotational
U(1) symmetry, but preserved time reversal invariance.
In this phase, the local symmetry of the $xx$ and $yy$ spin
correlations is changed by the
$\uparrow \uparrow \downarrow \downarrow$
and $\uparrow \downarrow \downarrow \uparrow$
structures, respectively. This reflects the sign-alternating character
of exchange couplings between neighboring spins at $|\delta| > J$
(see Eq.(\ref {eq:basic})).
The new spin correlations
(following the same power-law decay with distance as
that in the XY phase) induce a two-spin SN ordering of the system
described by the order parameter
$<vac|S^{x}_{n} S^{x}_{n+1} - S^{y}_{n} S^{y}_{n+1}|vac>
\sim (-1)^{n}~sgn~\delta$.
We also find a manifestation of Villain's
"order from disorder" \cite {Vil} - the effect of thermally
induced SN ordering of the spins
in the XY-like phase $(|\delta| < J)$.

In Sec.4 we extend consideration to the case of a finite
$\Delta$ to study effects of a weak staggered XY anisotropy ($|\delta| \ll J$)
in a Luttinger spin liquid.
Using bosonization method, we show that
 the low-energy properties of the system (2) are described
by a self-dual
field-theoretical model, similar to the one recently proposed
to describe two weakly coupled Luttinger chains \cite{KLN}.
 In the range
$-1/\sqrt{2} < \Delta < 1$,
 the off-diagonal perturbation $H_{\delta}$, tending to split
the velocities of fermionic excitations, is irrelevant
due to
the infrared catastrophe that eliminates single-fermion states from
the low-energy part of the spectrum. This behavior is an example of
Anderson's confinement \cite {conf} which has been extensively
discussed recently in connection with the two-chain problem.
In this regime, the system maintains basic properties of the gapless
Luttinger liquid, slightly modified by thermally induced SN order.
At $\Delta < -1/\sqrt{2}$, a translationally invariant, two-axis
anizotropy, generated in the second order in $\delta$, becomes a
relevant perturbation, and the system undergoes a Kosterlitz-Thouless
transition to a Neel-ordered state with polarization axis
$\hat{x}$ or $\hat{y}$.

By the $\delta \leftrightarrow J$ interchange symmetry, the above
picture also holds in the opposite limiting case
$|\delta| \gg J$. Here one finds Luttinger-liquid-like SN phases
and Kosterlitz-Thouless transitions to phases with
$\uparrow \uparrow \downarrow \downarrow$ ordering of the spins,
either along $\hat{x}-$ or $\hat{y}-$direction.

\section {Choice of the Model}

\par
Consider a bond-alternating S = 1/2 quantum XYZ spin chain given by the
Hamiltonian
\begin{equation}
H = \sum_{n=1}^{N} \sum_{\alpha = x,y,z} [J_{\alpha} + (-1)^{n} J_{\alpha}^{'}]
S^{\alpha}_{n} S^{\alpha}_{n+1}, \label {eq:stag}
\end{equation}
The model can be mapped onto two coupled quantum Ising chains by
a nonlocal unitary transformation,
previously used by Kohmoto, den Nijs and Kadanoff \cite {KNK}
in their study of the 2D Ashkin-Teller model. This transformation is
a special combination of
spin rotation and duality transformation \cite {Kom-Tas}.
Its basic steps are:
i) a global $\pi/2$ spin rotation, $\sigma_{n}^{y} \rightarrow
\sigma_{n}^{z}, ~\sigma_{n}^{z} \rightarrow -\sigma_{n}^{y}$; ii) duality
transformation
$ \sigma^{x}_{n} \rightarrow \sigma^{z}_{n-1/2} \sigma^{z}_{n+1/2}, ~
\sigma^{z}_{n} \sigma^{z}_{n+1} \rightarrow  \sigma^{x}_{n+1/2}$, where
$\{n+1/2\}$ are
the dual
lattice sites; iii) relabeling the sites, $ j \rightarrow \frac{1}{2} (j +
\frac {1}{4}) $,
and introduction of two kinds of the Pauli matrices,
$\sigma^{\alpha}_{j}$ and $\tau^{\alpha}_{j+1/2}$, defined on the two
sublattices of the dual
lattice; iv) duality transformation of the $\tau$-spins only. As a result, spin
operators
$S^{\alpha}_{n}$ of the staggered XYZ chain (2) are represented in terms of
$\sigma$ and $\tau$ matrices as products of order and disorder operators:
\begin{equation}
\begin{array}{clcr}
2S^{x}_{2j} = \sigma^{z}_{j} \tau^{z}_{j+1/2}, & 2S^{x}_{2j+1}
= \sigma^{z}_{j+1} \tau^{z}_{j+1/2}, \\
2S^{y}_{2j} = \sigma^{z}_{j+1/2} \tau^{z}_{j}, & 2S^{y}_{2j+1}
= \sigma^{z}_{j+1/2} \tau^{z}_{j+1}, \\
2S^{z}_{2j} = i\sigma^{z}_{j} \sigma^{z}_{j+1/2} \tau^{z}_{j} \tau^{z}_{j+1/2},
&
2S^{z}_{2j+1} = i\sigma^{z}_{j+1/2} \sigma^{z}_{j+1} \tau^{z}_{j+1/2}
\tau^{z}_{j+1},
\label{eq:KNK}
\end{array}
\end{equation}
where
$$ \sigma^{z}_{j+1/2} = \prod _{l=j+1}^{N/2}\sigma^{x}_{l},
{}~~~\tau^{z}_{j+1/2} = \prod _{l=1}^{j} \tau^{x}_{l}.
$$
The transformed Hamiltonian
$$
H = H_{\sigma} + H_{\tau} + H_{\sigma \tau}
$$
\begin{equation}
 = \sum_{j=1}^{N/2} [( A_{x} \sigma^{z}_{j} \sigma^{z}_{j+1} + B_{y}
\sigma^{x}_{j} )
+ ( A_{y} \tau^{z}_{j} \tau^{z}_{j+1} + B_{x} \tau^{x}_{j} )
- ( A_{z} \sigma^{z}_{j} \sigma^{z}_{j+1} \tau^{z}_{j} \tau^{z}_{j+1}
+ B_{z} \sigma^{x}_{j} \tau^{x}_{j} ) \label{eq:2chain}
\end{equation}
describes two coupled quantum Ising chains.
The constants $A_{\alpha} = (1/4)(J_{\alpha} + J_{\alpha}^{'}),$~
$B_{\alpha} = (1/4)(J_{\alpha} - J_{\alpha}^{'})$.
Following Ref.10, it can be shown that this Hamiltonian determines
the transfer matrix of the two-dimensional asymmetric Ashkin-Teller (AT) in the
highly
anisotropic, the so-called $\tau$-continuum limit. The asymmetric AT model
represents two
{\it different}
2D Ising models coupled by a four-spin interaction.
\par
To sort out those cases when the model (\ref{eq:2chain}) may show a critical
(gapless)
behavior, we shall
treat the interchain $\sigma \tau$ coupling as a weak perturbation and then
consider
conditions when at least one of the two quantum Ising chains is critical.
There are three such
possibilities.
\par
1) $H_{\sigma}$ is critical, while $H_{\tau}$ is not (or vice versa), implying
that
$ A_{x} = \pm B_{y},$$ A_{y} \neq \pm B_{x}.$
Using the Majorana fermion representation for each chain \cite {ID},
one finds one
massless
($H_{\sigma}$) and one massive ($H_{\tau}$) branch in the excitation spectrum.
The interchain interaction $H_{\sigma \tau}$ is irrelevant; its role is
exhausted
by renormalization of the mass and critical temperatute (or group velocity of
the massless
fermion). In this case, the critical behavior of the system belongs to the 2d
Ising
universality class.
\par
2)  $H_{\sigma}$ and $H_{\tau}$ are identical and critical; e.g.
$ A_{x} = B_{y} = A_{y} = B_{x}.$
If at the same time $ A_{z} = B_{z}$, the model (\ref{eq:stag}) becomes
equivalent to the
exactly solved
 U(1)-symmetric,
translationally invariant XXZ spin chain \cite {XXZ}, related to the six-vertex
model
\cite {Bax}.
The interchain
coupling $H_{\sigma \tau}$ is marginal: the system shows a Luttinger-liquid
critical
behaviour
with continuously varying (coupling-dependent) critical exponents.
\par
3) $H_{\sigma}$ and $H_{\tau}$ are both critical but {\it different}. This case
occurs when
\begin{equation}
A_{x} = \pm B_{y},~~  A_{y} = \pm B_{x}, ~~ |A_{x}| \neq |A_{y}|
\label{eq:cond}
\end{equation}
and corresponds to two massless Majorana fields with {\it different}
velocities.
This is a new, yet not considered possibility.
\par
Conditions (\ref{eq:cond}) give four possible distributions of the exchange
couplings
$ A_{x}, A_{y}, B_{x}$ and
$B_{y}$ on lattice bonds, each distribution breaking
both U(1) spin rotational symmetry and translational invariance (below we
indicate only the
$XY$-part of the corresponding Hamiltonians):
\begin{eqnarray*}
H_{1} &=& 2(A+B) \sum_{n} (S^{x}_{n} S^{x}_{n+1} + S^{y}_{n} S^{y}_{n+1})
+ 2(A-B) \sum_{n} (-1)^{n} (S^{x}_{n} S^{x}_{n+1} - S^{y}_{n} S^{y}_{n+1}),\ \\
H_{2} &=& 2(A-B) \sum_{n} (-1)^{n}
(S^{x}_{n} S^{x}_{n+1} + S^{y}_{n} S^{y}_{n+1})
+ 2(A+B) \sum_{n} (S^{x}_{n} S^{x}_{n+1} - S^{y}_{n} S^{y}_{n+1}),\ \\
\end{eqnarray*}
\begin{eqnarray*}
H_{3} &=& 2\sum_{n} [A - (-1)^{n} B]
(S^{x}_{n} S^{x}_{n+1} + S^{y}_{n} S^{y}_{n+1})
+ 2\sum_{n} [B + (-1)^{n} A] (S^{x}_{n} S^{x}_{n+1} - S^{y}_{n} S^{y}_{n+1}),\
\\
H_{4} &=& 2\sum_{n} [B + (-1)^{n} A]
(S^{x}_{n} S^{x}_{n+1} + S^{y}_{n} S^{y}_{n+1})
+ 2\sum_{n} [A - (-1)^{n} B] (S^{x}_{n} S^{x}_{n+1} - S^{y}_{n} S^{y}_{n+1}).
\end{eqnarray*}
All these cases are related to each other by nonlocal unitary transformations.
Transforming
$S^{y}_{n} \rightarrow (-1)^{n} S^{y}_{n},~~
S^{z}_{n} \rightarrow (-1)^{n} S^{z}_{n}$, one finds that
$$H_{2}(A,B;J_{z},J^{'}_{z}) \leftrightarrow H_{1}(A,B;-J_{z},-J^{'}_{z}), ~~
H_{4}(A,B;J_{z},J^{'}_{z}) \leftrightarrow H_{3}(A,B;-J_{z},-J^{'}_{z}).$$
On the other hand, under $S^{y}_{n} \rightarrow - S^{y}_{n}, ~~
S^{z}_{n} \rightarrow - S^{z}_{n}$ only at n = 4j + 1 and n = 4j + 2,
$$H_{3}(A,B;J_{z},J^{'}_{z}) \leftrightarrow H_{1}(A,B;-J_{z},-J^{'}_{z}).$$
In what follows, we shall concentrate on $H_1$. Choosing then $A_z = B_z$,
we arrive at the XXZ spin
chain with XY bond-alternating (staggered) anisotropy, given by
Eq.(\ref{eq:basic}).
It is equivalent to
the following Hamiltonian describing two critical
but different quantum Ising chains coupled by a self-dual interchain
interaction:
\begin{equation}
H = \frac{1}{4} \sum_{j=1}^{N/2} [(J - \delta) (\sigma^{z}_{j} \sigma^{z}_{j+1}
+ \sigma^{x}_{j} ) + (J + \delta) (\tau^{z}_{j} \tau^{z}_{j+1} + \tau^{x}_{j} )
- J \Delta (\sigma^{z}_{j} \sigma^{z}_{j+1} \tau^{z}_{j} \tau^{z}_{j+1}
+ \sigma^{x}_{j} \tau^{x}_{j} )]. \label{eq: 2sdchain}
\end{equation}
\par
The model (\ref{eq:basic}) has the following symmetry properties which will be
used below:
\par
1) Under a global $\pi/2$ rotation
$S^{x}_{n} \rightarrow S^{y}_{n}, ~~S^{y}_{n} \rightarrow -S^{x}_{n} $
\begin{equation}
H(J, \delta, J_{z}) \rightarrow H(J, -\delta, J_{z}). \label{eq: sym1}
\end{equation}
\par
2) Under a staggered transformation, $S^{x}_{n} \rightarrow (-1)^{n} S^{x}_{n},
{}~~
S^{y}_{n} \rightarrow (-1)^{n+1} S^{y}_{n}$
\begin{equation}
H(J, \delta, J_{z}) \rightarrow H(-J, \delta, J_{z}). \label{eq: sym2}
\end{equation}
\par
3) A nonlocal transformation to a $\pi/2$-twisted reference frame,
\begin{eqnarray}
S^{x}_{n} &\rightarrow& - S^{x}_{n}~~~~(n = 4j + 1, 4j + 2),\nonumber \\
S^{y}_{n} &\rightarrow& - S^{y}_{n}~~~~(n = 4j + 2, 4j + 3), \nonumber \\
S^{z}_{n} &\rightarrow& (-1)^{n} S^{z}_{n} \label {eq: twstr}
\end{eqnarray}
leads to the important symmetry under interchange of $J$ and $\delta$:
\begin{equation}
H(J, \delta, J_{z}) \rightarrow H(\delta, J, - J_{z}). \label{eq: sym3}
\end{equation}

\section {XY chain with Bond-Alternating Anisotropy}

In this section we set $J_{z} = 0$ and consider properties of the XY chain with
a bond-
alternating anisotropy:
\begin{equation}
H = H_{0} + H_{\delta}
= \sum_{n=1}^{N} [ J (S^{x}_{n} S^{x}_{n+1} + S^{y}_{n} S^{y}_{n+1}) -
\delta (-1)^{n} (S^{x}_{n} S^{x}_{n+1} - S^{y}_{n} S^{y}_{n+1})]. \label{eq:
altXY}
\end{equation}
Due to the symmetry property (\ref{eq: sym2}), we can choose without loss of
generality
$J > 0$.
The Jordan-Wigner (JW) transformation
\begin{equation}
S^{+}_{n} = a^{+}_{n} \exp (-i \pi \sum_{j=1}^{n-1} a^{+}_{j} a_{j}), ~~
S^{z}_{n} = :a^{+}_{n} a_{n}: \equiv a^{+}_{n} a_{n} - \frac{1}{2} \label{eq:
JW}
\end{equation}
represents Hamiltonian (\ref{eq: altXY}) as a quadratic form of spinless
fermion operators
\begin{equation}
H = \frac{1}{2} \sum_{n}[J (a^{+}_{n} a_{n+1} + a^{+}_{n+1} a_{n} ) -
\delta (-1)^{n} (a^{+}_{n} a^{+}_{n+1} + a_{n+1} a_{n} ) ], \label{eq: SF}
\end{equation}
describing a tight-binding half-filled band and a Cooper pairing of the
fermions on
neighboring sites with a sign-alternating amplitude.
The latter circumstance leads to commutativity of $H_{0}$ and $H_{\delta}$
which,
in turn, indicates the absence of a gap in the spectrum
at arbitrary values of $J$ and $\delta$.
This is in agreement with the equivalent representation of
model (\ref {eq:basic}) in terms of two decoupled quantum Ising chains {\it at
criticality}, as given by Eq.(\ref {eq: 2sdchain}) at $\Delta = 0$.
Representation (\ref {eq: SF}) implies that the excitation spectrum consists of
two
branches which, in the continuum limit,
correspond to two massless Majorana fields with group velocities
$v_{\pm} = |J \pm \delta| $ (we set the lattice constant $a=1$).
\par
To understand the effect of the staggered XY anisotropy on the ground state
properties
of the model  (\ref {eq: SF}), we rewrite the Hamiltonian in momentum
representation
\begin{equation}
H = H_{0} + H_{\delta}
= \sum_{|k|<\pi} [\epsilon (k) a^{+}_{k} a_{k} + \frac{1}{2} \Delta (k)
(a^{+}_{k} a^{+}_{\pi-k} + h.c.)], \label{eq: BCS}
\end{equation}
where
$$\epsilon (k) = - \epsilon (k+\pi) = J \cos k, ~~~
\Delta (k) = - \Delta (k+\pi) = \delta \cos k. $$
We see that $ H_{\delta}$ describes Cooper pairing of JW fermions
with total momentum equal $\pi$.
Mapping Hamiltonian (\ref {eq: BCS}) onto reduced Brillouin zone,
$|q| < \pi/2$
$$
H = \sum_{|q| < \pi/2} [\epsilon (q) (a^{+}_{q} a_{q} - a^{+}_{\pi-q}
a_{\pi-q})
+ \Delta (q) (a^{+}_{q} a^{+}_{\pi-q} + a_{\pi-q} a_{q})],
$$
and introducing a Nambu 2-spinor
$$
\psi_{q} = \left( \begin{array}{c}
                   a_{q} \\
                   a^{+}_{\pi-q} \end{array} \right),
$$
one rewrites $H$ in the form
\begin{equation}
H = E_{0} + \sum_{|q| < \pi/2} \psi^{+}_{q}~ [\epsilon (q) + \Delta (q)
\tau^{1}] ~\psi_{q},
\label {eq: Nambu}
\end{equation}
where $\tau^{\alpha}$ are the Pauli matrices, and
\begin{equation}
 E_{0} = - \sum_{q} \epsilon (q) = -\frac {NJ}{\pi} \label {eq: vacen}
\end{equation}
is the ground state energy of $H_{0}$.
\par
Eq.(\ref {eq: Nambu}) explains the origin of commutativity of $H_{0}$ and
$H_{\delta}$. The
Hamiltonian of the "pure" XY model,  $H_{0}$, with global U(1) spin rotation
symmetry in real space, posesses a local SU(2) pseudospin symmetry
in momentum space \cite {SCZ}. Generators of this SU(2) group are
\cite{foot}
\begin{equation}
J^{\alpha}_{q} = \frac{1}{2} \psi_{q}^{+} \tau^{\alpha} \psi_{q}
{}~~(\alpha = x,y,z). \label {eq: gen}
\end{equation}
For a given $q$, $\Delta (q)$ then appears as a "magnetic" field oriented along
the $x$-axis in the local pseudospin space.
\par
Diagonalization of $H$ reduces to a $\pi/2$-rotation
$\psi_{q} \rightarrow \exp [- (\pi/2) J^{2}_{q}]~ \psi_{q},$
under which
\begin{eqnarray}
H &=& E_{0} + H_{exc},\nonumber\\
 H_{exc} &=& \sum_{|q| < \pi/2} [E_{+}(q) \alpha^{+}_{q} \alpha_{q} +
E_{-}(q) \beta^{+}_{q} \beta_{q}], \label {eq: diag}
\end{eqnarray}
where new fermion operators
\begin{equation}
\alpha_{q} = \frac{1}{\sqrt{2}} (a_{q} + a^{+}_{\pi -q}), ~~~
\beta_{q} = \frac{1}{\sqrt{2}} (- a_{q} + a^{+}_{\pi -q}) \label {eq: Bogol}
\end{equation}
and
\begin{equation}
E_{\pm}(q) = (J \pm \delta) \cos q   \label {eq: spec}
\end{equation}
\par
At $|\delta| < J$ $E_{\pm}(q) \geq 0$ at all $|q| < \pi/2$. Therefore in
Eq.(\ref {eq: diag})
$H_{exc}$ is positive definite and represents the excitation energy of the
system,
$E_{\pm}(q)$ thus being energies of the single-fermion excitations.
The ground state $|vac>$ is defined via relations
$\alpha_{q}|vac> = \beta_{q}|vac> = 0$ and coincides with that of $H_{0}$.
The local pseudospin SU(2) symmetry
is unbroken: $J^{\alpha}_{q} |vac> = 0$.
\par
Thus, at $|\delta| < J$, the alternating XY anisotropy
has no effect on the ground state of the system.
This can be understood from the following simple arguments.
Start with the ground state of the pure XY model $H_{0}$,
$$|vac> = \prod_{\pi/2<|k|<\pi} a^{+}_{k} |0> $$
and then consider $H_{\delta}$ as a perturbation. Note that if a single-fermion
state $|k>$ is
occupied, the state $|\pi-k>$ is necessarily empty, or vice versa. As a result,
the Cooper
pairing
with total momentum $\pi$ cannot occur in the ground state, if $\delta$ is
small
enough; so the ground
state is that of the XY model. This means that there is no off-diagonal
long-range
order (ODLRO) in the ground state
\begin{eqnarray}
<vac|a^{+}_{k} a^{+}_{\pi-k}|vac> &=& 0,\nonumber\\
<vac|S^{x}_{n} S^{x}_{n+1} - S^{y}_{n} S^{y}_{n+1}|vac> &=& 0, ~~(|\delta| <
J),
\label{eq: ODLRO}
\end{eqnarray}
implying that spin rotational U(1) symmetry and translational invariance remain
intact.
However, perturbation $H_{\delta}$ reveals at the level of excitations.
Consider
one-particle and one-hole excitations
$|q>_{p} = a^{+}_{q} |vac>, ~~ |\pi - q>_{h} = a_{\pi - q} |vac> $
with $|q| < \pi/2$. At $\delta = 0$ these states are degenerate:
$ E_{p}(q) = E_{h}(\pi - q) = J \cos q \geq 0 ~~ (J > 0) $.
The degeneracy is removed by
$H_{\delta}$, but only at $ q \neq \pm k_{F}$, since the pairing
amplitude $\Delta (q)$ has nodes
at the two Fermi points. Therefore, a gap does not open; instead the doubly
degenerate
spectrum splits into two gapless branches with different Fermi velocities.
\par
The above picture persists at all $ |\delta| < J$. However,
at $ |\delta| > J$ one of the two excitation energies (\ref {eq: spec}) becomes
negative,
and the
ground state changes.
Since $[H_{0}, H_{\delta}] = 0$, this simply occurs due to level crossing: one
of the
former
excited states becomes the ground state. As a result, the symmetry of the
vacuum is
changed. Choosing, {\it e.g.}, $\delta > J > 0$, and making a particle-hole
transformation $\beta_{q} \rightarrow \bar{\beta}^{+}_{q}$, one finds that the
ground state is now determined by $\alpha_{q}|vac> = \bar{\beta}_{q}|vac> = 0$,
the ground state
energy and Hamiltonian of excitations being changed, respectively, by
\begin{equation}
E_{0} = - \sum_{q} \Delta (q) = - \frac{N |\delta|}{\pi}, \label{eq: vacen1}
\end{equation}
\begin{equation}
 H_{exc} = \sum_{|q| < \pi} [E_{+}(q) \alpha^{+}_{q} \alpha_{q} +
E_{-}(q) \bar{\beta}^{+}_{q} \bar{\beta}_{q}] \label{eq: diag1}
\end{equation}
with
\begin{equation}
E_{\pm}(q) = (\delta \pm J) \cos q \geq 0. \label{eq: spec1}
\end{equation}
\par
Notice that, at $ |\delta| > J$, $H_{0}$ does not contribute to the ground
state energy:
\begin{equation}
<vac|S^{x}_{n} S^{x}_{n+1} + S^{y}_{n} S^{y}_{n+1}|vac> = 0.
\end{equation}
The local pseudospin SU(2) symmetry is now broken down to U(1),
$ <vac|J^{1}_{q}|vac> = - 1/2 $, implying that the ground state
is characterized by a nonzero ODLRO which breaks global U(1) and translational
symmetries:
\begin{equation}
- \frac{1}{N} \frac{\partial E_{0}}{\partial \delta} =
\frac{1}{N} \sum_{n} (-1)^{n} <vac|S^{x}_{n} S^{x}_{n+1} - S^{y}_{n}
S^{y}_{n+1}|vac> =
\frac{1}{\pi}
\theta (\delta^{2} - J^{2})~
sgn~\delta, \label {eq: ODLRO1}
\end{equation}
where $\theta (x)$ is the step function.
Note that all results concerning the case  $ |\delta| > J$ could have also been
obtained by
using the $\delta \leftrightarrow J$ interchange symmetry, Eq. (\ref {eq:
sym2}).
\par
Using the two-chain representation (\ref {eq:KNK}) of the spin operators, one
finds
that, at all values of
$J$ and $\delta$, the local magnetization of the model (\ref {eq: altXY})
vanishes:
$<S^{\alpha}_{n}> = 0$.
This simply follows from the fact that, since the $\sigma$ and $\tau$ quantum
Ising
chains are
decoupled and self-dual (critical), the $Z_{2} \times Z_{2}$ symmetry related
to
transformations
$\sigma^{z}_{j} \rightarrow -\sigma^{z}_{j},
{}~~\tau^{z}_{j} \rightarrow -\tau^{z}_{j}$ remains unbroken. Absence of local
magnetization
means that time reversal symmetry is always preserved. Preservation of time
reversal symmetry
together
with breakdown of spin rotational U(1) symmetry allows to conclude that, at
$|\delta| > J$,
the
ground state
represents a {\it spin-nematic (SN) phase} \cite {A-G}, with the order
parameter given by
symmetric tensor (\ref {eq: ODLRO1}).
The SN state is doubly ($Z_{2}$) degenerate; the two SN phases at $\delta > J$
and
$\delta < -J$
transform to each other under transformation (\ref {eq: sym1}).

\par
Let us now consider spin correlations in the ground state of the XY-like
($|\delta| < J$) and SN ($|\delta| > J$) phases.
According to the two-chain representation (\ref {eq: 2sdchain}) of the
Hamiltonian
and multiplicative $\sigma \tau$ form (\ref {eq:KNK}) of the
spin operators, the correlation functions factorize into products of
independent contributions of each quantum Ising chain. At large space-time
separations, the $\sigma$ and $(\tau)$ contributions are determined by gapless
excitations,
described in terms of (right- and left-moving) Majorana fermions with
velocities $v_{+}$ and $v_{-}$, respectively.
We shall first consider the case $|\delta| < J$.
Using the Majorana fermion representation of the quantum Ising chain
\cite {ID}
and passing to a continuum limit, one can show that
the longitudinal $(zz)$ spin correlation
function is bilinear in the Majorana single-fermion Green's functions:
$$<S^{z}_{n}(t) S^{z}_{m}(0)> \simeq - \frac{1}{4\pi^{2}}
\left ( \frac{1}{x + v_{+}t} - \frac {(-1)^{n-m}}{x - v_{+}t} \right )
\left ( \frac{1}{x + v_{-}t} - \frac {(-1)^{n-m}}{x - v_{-}t} \right )$$
\begin{equation}
= -\frac{1}{2\pi^{2}} \frac{[1 - (-1)^{n-m}]x^{2} + [1 + (-1)^{n-m}]
v_{+}v_{-}t^{2} }
{(x^{2} - v^{2}_{+}t^{2})(x^{2} - v^{2}_{-}t^{2})}, \label {eq:zz}
\end{equation}
where $x = n - m$.
This result can also be obtained by using the JW representation (\ref {eq: JW})
for
$S^{z}_{n}$
together with Bogoliubov transformation (\ref {eq: Bogol}) to quasiparticle
operators
$\alpha_{q}$
and $\beta_{q}$. The transverse spin correlation function is a product of two
Ising correlation functions:
\begin{equation}
<S^{x}_{n}(t) S^{x}_{m}(0)>~ = ~<S^{y}_{n}(t) S^{y}_{m}(0)>~
\sim ~- (-1)^{n-m} \frac {1}{(x^{2} - v^{2}_{+}t^{2})^{1/8}}
\frac {1}{(x^{2} - v^{2}_{-}t^{2})^{1/8}}. \label {eq:xx}
\end{equation}
At equal times, formulas (\ref{eq:zz}) and (\ref{eq:xx}) coincide with known
results
for the "pure"
XY model \cite {LSM} $(\delta = 0)$.
\par
Now we consider the case $|\delta| > J$. For simplicity, we shall restrict
ourselves by equal time correlations.
As follows from the $\delta \leftrightarrow J$ symmetry transformations
(\ref {eq: twstr}), (\ref {eq: sym3}),
in the SN phases
the antiferromagnetic $zz$-correlations (\ref{eq:zz}) transform to the
ferromagnetic
ones:
\begin{equation}
<S^{z}_{n} S^{z}_{m}>_{vac} =  \frac{1}{2\pi^{2}}
\frac{[1 - (-1)^{n-m}]}{|n-m|^{2}} \label {eq: ferrozz}
\end{equation}
In the XY-like phase, the power-law decay (\ref{eq:zz}) of the
antiferromagnetic
$zz$-correlations
leads to the well known singular response of the system to a staggered magnetic
field along the
$z$-axis, logarithmically divergent at $T \rightarrow 0$.
(In terms of the JW fermions, this property appears as a logarithmic
charge-density-wave instability of a noninteracting 1D Fermi system).
Similarly,
in the SN phases, ferromagnetic $zz$-correlations (\ref {eq: ferrozz}) give
rise to
a logarithmically divergent response to a homogeneous
magnetic field. The ferromagnetic and antiferromagnetic
susceptibilities are finite at $|\delta| < J$ and $|\delta| > J$, respectively,
$$
\chi_{F} \sim \frac{1}{\delta} \ln \left( \frac{J + \delta}{J - \delta}
\right),
{}~~~(|\delta| < J),
$$
$$
\chi_{AF} \sim \frac{1}{J}  \ln \left( \frac{\delta + J}{\delta - J} \right),
{}~~~(|\delta| > J).
$$
At the boundaries $\delta = \pm J$ between the XY-like and SN phases,
the model shows both ferromagnetic and antiferromagnetic instabilities.
\par
Applying the $\delta \leftrightarrow J$ transformations (\ref {eq: twstr}) to
(\ref{eq:xx}), we find
that, at $|\delta| > J$ , the transverse spin correlations show the same
power-law
decay as in the XY-like phase, but are characterized by a different type
of local ordering:
\begin{equation}
<S^{x}_{2j} S^{x}_{2j+2l}> \sim \frac{(-1)^{l}}{|2l|^{1/2}}, ~~~
<S^{x}_{2j} S^{x}_{2j+2l+1}> \sim \frac{(-1)^{l}}{|2l + 1|^{1/2}}, \label {eq:
xx1}
\end{equation}
\begin{equation}
<S^{y}_{2j} S^{y}_{2j+2l}> \sim \frac{(-1)^{l}}{|2l|^{1/2}}, ~~~
<S^{y}_{2j} S^{y}_{2j+2l+1}> \sim \frac{(-1)^{l+1}}{|2l + 1|^{1/2}}. \label
{eq: xx2}
\end{equation}
Eqs.(\ref {eq: xx1}) and (\ref {eq: xx2}) describe
$\uparrow \uparrow \downarrow \downarrow$ and
$\uparrow \downarrow \downarrow \uparrow$ periodic structures
for the $xx$ and $yy$ spin correlations, respectively.
The origin of these correlations is easily seen from (\ref {eq: altXY}).
At $\delta > J$ the
$x$-components of neighboring spins are coupled ferromagnetically
on even bonds $<2j, 2j+1>$ and antiferromagnetically on odd bonds
$<2j-1, 2j>$, whereas for the $y$-components the picture is opposite
(at $\delta < -J$ one has simply to interchange $x$- and $y$-components).
\par
The transverse spin correlations (\ref {eq: xx1}), (\ref {eq: xx2}) locally
describe
a classical ground
state of Hamiltonian $H_{\delta}$. Such a state represents a periodic
structure,
with a unit cell consisting of four spins ${\bf S}_{m} = \hat{x} \cos
\varphi_{m}
+ \hat{y} \sin \varphi_{m}, ~m = 1,2,3,4$, where angles $\varphi_{m}$ are given
by
$\varphi_{1} = \varphi,$ $\varphi_{2} = \pi - \varphi,$ $\varphi_{3} = \pi +
\varphi,$
$\varphi_{4} = -\varphi$ at $\delta > 0$, and
$\varphi_{1} = \varphi,$ $\varphi_{2} = -\varphi$, $\varphi_{3} = \pi +
\varphi,$
$\varphi_{4} = \pi - \varphi$ at $\delta < 0$,
$\varphi$ being an arbitrary angle. These configurations
can be viewed as two interpenetrating antiferromagnetic sublattices whose
staggered magnetizations transform to each other under reflection about the
$y$-axis ($\delta > 0$), or $x$-axis ($\delta < 0$),
as required by the symmetry of $H_{\delta}$.
Apparently, these states can be obtained from a classical Neel ground state of
the
XY model $H_{0}$ by means of transformations (\ref {eq: sym3}) and (\ref {eq:
sym1}).
Furthermore,
the nonlocal transformation (\ref {eq: twstr}) maps the continuous symmetry of
$H_{0}$ under uniform
U(1) spin rotations onto the symmetry of $H_{\delta}$ under continuous
rotations of the two sublattices by the same angle but in opposite directions.
As a result, the classical ground state of  $H_{\delta}$ is continuously
degenerate under varying the angle $\varphi$.
\par
In the quantum case of interest,
the corresponding gapless "phason" mode destroys single-spin long-range order
precisely in the same way as it occurs in the XY model. However, despite the
loss of local magnetization, the survival of transverse correlations (\ref {eq:
xx1}),
(\ref {eq: xx2})
gives rize to SN ordering of the system, described by two-spin
tensor (\ref {eq: ODLRO1}). This is similar to the picture described by
Chandra,
Coleman and Larkin \cite {CCL1} for 2d frustrated
antiferromagnets, where short-range correlations with local twist structure
lead to the formation of a SN state. In the latter case, the order
parameter is a parity-breaking antisymmetric tensor
${\bf T}(x,x^{'}) = < {\bf S}(x) \times {\bf S}(x^{'})>$, corresponding to
the p-type SN phase \cite {A-G}. The difference between this and our cases is
the
absence of local twist structure
in the ground state of our model at $|\delta| > J$, (${\bf T}(x,x^{'}) = 0$).
Instead we have the above described local spin structure generated by the
staggered XY anisotropy, which results in a SN order parameter (\ref {eq:
ODLRO1}) with
a symmetric
tensor form.
\par
To conclude this section, let us briefly discuss the role of finite
temperature.
As it was already pointed out, the staggered XY anisotropy does not affect the
ground state
of the model at $\delta < J$, but reveals in splitting the excitation spectrum
in two
gapless branches with different Fermi velocities. One of interesting
manifestations of
this splitting is the appearance of the temperature-induced SN long-range order
in the XY phase.
Using the diagonalized form (\ref {eq: diag}) of the Hamiltonian, one finds
that
\begin{eqnarray*}
<S^{x}_{2j} S^{x}_{2j+1}> = <S^{y}_{2j-1} S^{y}_{2j}> =
- \frac{1}{2\pi} + \frac{1}{J - \delta} \frac{{\it E}_{-}}{N},\ \\
<S^{x}_{2j-1} S^{x}_{2j}> = <S^{y}_{2j} S^{y}_{2j+1}> =
- \frac{1}{2\pi} + \frac{1}{J + \delta} \frac{{\it E}_{+}}{N}.
\end{eqnarray*}
Here
$ {\it E}_{\pm}$
are the average thermal energies of the $\alpha$- and $\beta$-quasiparticles.
At low
temperatures, $T \ll J \pm \delta$,
${\it E}_{\pm}/N = (\pi/24) (T^{2}/(J \pm \delta))$.
Assuming for simplicity that $\delta \ll J$ one finds
$$
<S^{x}_{2j} S^{x}_{2j+1}> = <S^{y}_{2j-1} S^{y}_{2j}> =
\left( -\frac{1}{2\pi} + \frac{\pi T^{2}}{24 J^{2}} \right) +
\frac{\pi}{12} \left( \frac{\delta}{J} \right) \left( \frac{T}{J}
\right)^{2},$$
$$<S^{x}_{2j-1} S^{x}_{2j}> = <S^{y}_{2j} S^{y}_{2j+1}> =
\left( -\frac{1}{2\pi} + \frac{\pi T^{2}}{24 J^{2}} \right) -
\frac{\pi}{12} \left( \frac{\delta}{J} \right) \left( \frac{T}{J} \right)^{2}.
$$
The first term in the right-hand sides of these equations describes
antiferromagnetic
correlations
in the XY model, slightly suppressed by thermal fluctuations. The second terms
with opposite
signs represent a combined effect of the temperature and XY bond-alternating
anisotropy.
These terms have the symmetry of the above described transverse spin
correlations in the
ground state of the SN phase
and give rise to a finite value of the SN order parameter at $|\delta| < J$
\begin{equation}
<S^{x}_{n} S^{x}_{n+1} - S^{y}_{n} S^{y}_{n+1}> = (-1)^{n}~
\left( \frac{\pi \delta}{6 J} \right) \left( \frac{T}{J} \right)^{2} \label
{eq: temp1}
\end{equation}
\par
The phenomenon we are dealing with here is an example of Villain's "order from
disorder"
\cite{Vil}.
Its manifestations in 2d frustrated magnets have been recently described in
Refs.17,18.
The $\delta$-anisotropy cannot influence the ground state at $\delta < J$;
however, it
lowers thermal excitation energy of the system by splitting the spectrum {\it
and}
repopulating
the excited quasiparticles. This leads to the generation of new spin
correlations, absent in
the ground state, indicating the onset of SN order from thermal disorder.
\par
On increasing $|\delta|$, one of the two modes becomes very
soft and nearly completely populated, when $|J - |\delta|| \ll T \ll J +
|\delta|$.
In this temperature range,
the $xx$ spin correlations on even (odd) bonds
and $yy$ correlations on odd (even) bonds get suppressed,
and the amplitude of the SN order parameter is close to $1/2 \pi$.
At $|\delta| \rightarrow J$,
the SN long-range order eventually penetrates into the ground state.

\section {XXZ Chain with a Weak Bond-Alternating XY Anisotropy}
In this section, we extend the above considerations to the model (\ref
{eq:basic}) with a
finite $J_{z}$ and
study effects of a weak bond-alternating XY anisotropy in the disordered
gapless phase of the
XXZ chain, assuming that
$|\delta| \ll J,$ $|\Delta| < 1.$
The question we address here is: How off-diagonal perturbation $H_{\delta}$,
tending to split velocities of fermionic excitations, reveals in a Luttinger
spin liquid
\cite {LP,Hald-LL}, where
orthogonality catastrophe suppresses single-fermion states in the low-energy
part of
the spectrum.
Notice that the $\delta \leftrightarrow J$ symmetry (\ref {eq: sym3}) allows to
translate
results of
this section to the case of very strong $\delta$-anisotropy,
$|\delta| \gg J$.
\par
Being interested in the infrared properties of the model, we pass to a
continuum description
developed for the XXZ spin chain by Luther and Peschel \cite {LP}. Linearizing
the spectrum
of the JW fermions in the vicinity of
two Fermi points $\pm k_{F} = \pm \pi/2$ and introducing two Fermi fields
corresponding to
right-moving ($\psi_{1}$) and left-moving ($\psi_{2}$) particles
\begin{equation}
a_{n} \rightarrow (-i)^{n} \psi_{1}(x) + i^{n} \psi_{2}(x), \label {eq: cont}
\end{equation}
one makes use of Abelian bosonization \cite {LP,Shan}
\begin{eqnarray}
:\psi^{+}_{1,2}(x) \psi_{1,2}(x):~= \frac{1}{\sqrt{\pi}} \partial_{x}
\varphi_{1,2} (x)
\nonumber\\
\psi_{1,2}(x) \rightarrow \frac {1}{\sqrt{2 \pi \alpha}}
\exp [\pm i \sqrt{4 \pi}~ \varphi_{1,2} (x)],  \label {eq:fb}
\end{eqnarray}
$$
\varphi_{1,2}(x) = \frac{1}{2} \left( \varphi(x) \mp \int^{x} dy~ \Pi(y)
\right),
$$
to describe the low-energy spin excitations in terms of
a scalar field theory
\begin{equation}
H_{XXZ} = \frac{1}{2} v_{F} \int dx [ :\Pi^{2}(x): + ( 1 + \frac{4 \Delta}{\pi}
)
:(\partial_{x} \varphi(x))^{2}:] + H_{U}.
\end{equation}
Here $\varphi(x)$ and $\Pi(x)$ are the scalar field and its conjugate momentum,
respectively.
The term
\begin{equation}
H_{U} \sim \Delta \int dx \cos (\sqrt{16 \pi} \varphi (x)) \label {eq: Umkl}
\end{equation}
originates from Umklapp scattering of the JW fermions \cite {Hald-LL,Umkl}.
At $\Delta > 1$, these processes
drive the system
to a strong-coupling, massive Neel phase, but are irrelevant in the disordered
phase,
$|\Delta| < 1$. Using continuum representation (\ref {eq: cont}), the staggered
XY
anisotropy $H_{\delta}$
transforms to a pairing term
\begin{equation}
H_{\delta} = - \frac{i \delta}{2} \int dx [ :\psi^{+}_{1}(x) \psi^{+}_{1}(x +
a): -
:\psi^{+}_{2}(x) \psi^{+}_{2}(x + a): - h.c.] \label {eq: dcont}
\end{equation}
Bosonizing (\ref {eq: dcont}) by means of (\ref {eq:fb})
and performing canonical transformation of the field and
momentum
\begin{equation}
\varphi (x) \rightarrow \frac{\beta}{\sqrt{4 \pi}} \phi (x), ~~
\Pi(x) \rightarrow \frac{\tilde{\beta}}{\sqrt{4 \pi}} P(x), ~~ \beta
\tilde{\beta} = 4 \pi,
\label {eq: canon}
\end{equation}
one arrives at the following self-dual field-theoretical model
\begin{equation}
H = \int dx [ \frac{u}{2} \left( :P^{2}(x): + :(\partial_{x} \phi (x))^{2}:
\right)
+ \frac{\delta}{\pi \alpha} \cos \beta \phi(x) \cos \tilde{\beta}
\tilde{\phi}(x) ].
\label {eq: sdmod}
\end{equation}
Here $u$ is the renormalized velocity, $\alpha$ is a cutoff parameter, and
$\tilde{\phi}(x)$
is a field dual to ${\phi}(x)$, defined as $\partial_{x} \tilde{\phi}(x) =
P(x)$.
The correct parametrization of $\beta$ is provided by the Bethe-ansatz solution
of the
XXZ model
\cite {XXZ}:
\begin{equation}
\frac{\beta^{2}}{2\pi} = \left( 1 - \frac{\theta}{\pi} \right)^{-1}, ~~~\cos
\theta = \Delta
\label {eq: par}
\end{equation}
\par
Amazingly, the model (\ref {eq: sdmod}) is similar to the one recently proposed
to describe
two weakly coupled spinless Luttinger chains \cite {KLN}.
The difference between the two models stems from different values of the
conformal spin
$S_c = \beta \tilde{\beta}/ 2\pi$ of the perturbation: $S_c = 1$ for the
two-chain model, and
$S_c = 2$ for the present one.
 As a result, the two-chain and
present models belong to different universality classes.
It was shown in Ref.8 that the self-dual theory (\ref {eq: sdmod}) can be
equivalently reformulated
in terms of a 2d Coulomb gas of
charge-monopole composites. The latter is given by the grand partition
function:
\begin{equation}
Z/Z_{0} = \sum_{n=0}^{\infty} \frac{z^{2n}_{\perp}} {(2n)!} \int
\prod_{j=1}^{2n} \frac {d^{2} {\bf x}_{j}}
{\alpha^{2}}~ \sum_{\sigma,\tilde{\sigma}}
\exp\left( 2K \sum_{i<j} \sigma_{i} \sigma_{j} l_{ij} + 2\tilde{K} \sum_{i<j}
\tilde{\sigma_{i}}
\tilde{\sigma_{j}} l_{ij} + 2i \sum_{i\neq j} \sigma_{i} \tilde{\sigma_{j}}
\varphi_{ij}
\right)
\label {eq: CG}
\end{equation}
where $Z_{0}$ is the partition function of the unperturbed, Gaussian part of
the model.
Eq.(\ref {eq: CG}) descibes a 2d system of classical particles with coordinates
${\bf x}_{j} = (u \tau_{j}, x_{j}),~
(0 < \tau_{j} < 1/T)$, $z = \delta \alpha /\pi u$ being the fugacity.
Each particle carries an "electric" charge $\sigma_{j} = \pm 1$ and
"magnetic" charge, or monopole, $\tilde{\sigma}_{j} = \pm 1$. The
four-component system of
charge-monopole composites is neutral with respect to each kind of charges. The
charges
and monopoles
interact with logarithmic potential $l_{ij} =  ln (|{\bf x}_{i} - {\bf x}_{j}|/
{\alpha})$
amongst themselves, and there is also a statistical Aharonov-Bohm phase
$\varphi_{ij} = \arctan ((x_{i} - x_{j})/u(\tau_{i} - \tau_{j}))$ which couples
the charges
and
monopoles belonging to different composites.
The parameters $ K = \beta^{2}/4 \pi$ and $ \tilde{K} = 1/K$
are naturally interpreted as inverse dimensionless temperatures for the charges
and
monopoles,
respectively.
\par
The perturbation in (\ref {eq: sdmod}) has critical dimension
$ D =  K + \tilde{K} \geq 2 $. So,
in the XY-like, gapless phase of the XXZ chain, the staggered XY anisotropy is
marginal at the
XY point $\Delta = 0$ and irrelevant at $\Delta \neq 0$. The renormalized
amplitude
$\delta_{eff}(T)$, and hence the difference between the velocities of two
Majorana
fermionic excitations,
scales down to zero according to the power law
$\delta_{eff}(T) \sim \delta (\alpha T/u)^{D - 2}$. This is analogous to the
confinement regime in the two-chain model \cite {conf}, in which
Luttinger-liquid
effects suppress single-particle
interchain hopping in the infrared limit \cite {KLN,Yak,NLK}.

At finite temperatures, thermally induced SN
ordering will take place. However,
in the Luttinger spin-liquid regime $(\Delta \neq 0)$ , the SN order parameter
increases with the temperature more slowly than
in a noninteracting Fermi gas, Eq.(\ref {eq: temp1}).
Using Coulomb gas representation (\ref {eq: CG}) and
calculating the second-order correction to the free energy of the XXZ model,
one finds that
\begin{equation}
(-1)^{n} <S^{x}_{n} S^{x}_{n+1} - S^{y}_{n} S^{y}_{n+1}> =
- \frac{1}{L} \frac{\partial F}{\partial \delta} \sim
\left( \frac{\delta \alpha}{u} \right) \left( \frac {T \alpha}{u} \right)^{2D -
2}
\label {eq: tempLut}
\end{equation}
The nonuniversal power-law temperature dependence of the SN order parameter
reflects
the well-known infrared catastrophe in a 1D Luttinger liquid that eliminates
single-fermion
states from the low-energy part of the spectrum. The r.h.s. of (\ref {eq:
tempLut}) should
then be understood as
$ \sim T^{2} [ N_{LL} (T)]^{2}$, where
$N_{LL}(\omega) \simeq N_{0} (|\omega| \alpha / u)^{D-2} $
is the single-fermion density of states vanishing in the zero-energy limit.
\par
The suppression of the XY anisotropy in the lowest order in $\delta$ does not
really mean that $H_{\delta}$ is a totally irrelevant pertubation. The usual
criterium of
relevance,
based on the comparison of the critical dimension of a pertubation with
space-time
dimension 2,
is not applicable here, since the pertubation has a nonzero conformal spin
\cite{spin}.
This is seen from the Coulomb gas representation (\ref {eq: CG}). The
suppression of the XY
staggered anisotropy
in the first order in $\delta$ originates from pairing of the charge-monopole
composites with
zero total "electric" and "magnetic" charges. On the other hand, binding of
composites
in pairs with
total "magnetic" or "electric" charge $\pm 2$ gives rise to two operators
\begin{equation}
O(x) = \cos 2 \beta \phi (x), ~~~~~~ \tilde{O}(x) = \cos 2 \tilde{\beta}
\tilde{\phi} (x)
\label {eq: op}
\end{equation}
which are absent in the original model (\ref {eq: sdmod}), but are generated
upon
renormalization in the
effective Hamiltonian in the
second order in $\delta$ \cite {KLN,Yak,NLK}.
The operators (\ref {eq: op}) have zero conformal spin; so their relevance can
be
established by the
usual criterium.
\par
Clearly, $O(x)$ is the Umklapp operator (see (\ref {eq: Umkl})), relevant only
at $|\Delta| > 1$.  $\tilde{O}(x)$ is a new operator whose critical dimension
is
$4 \tilde{K}$.
It becomes marginal at the point  $\tilde{K} = 1/2$, or equivalently,
$\Delta = - \frac{1}{\sqrt{2}}$. Therefore, in the region
$- \frac{1}{\sqrt{2}} < \Delta < 1$,
where both operators $O(x)$ and $\tilde{O}(x)$ are irrelevant,
the model (\ref {eq: sdmod}) preserves basic features
of the gapless
Luttinger-liquid phase of the XXZ chain, slightly modified by thermally induced
SN ordering
(\ref {eq: tempLut}).
This is a pure realization of Anderson's confinement \cite {conf},
i.e. stability of the Luttinger
liquid against single-fermion off-diagonal perturbations, as opposed to the
case of
two coupled Luttinger chains, where the generation of two-particle interchain
correlations
drives inevitably the system away form the Luttinger-liquid fixed point at
arbitrary
in-chain interaction
(see, {\it e.g.} Ref.24 and references therein).
\par
The point $\Delta = - \frac{1}{\sqrt{2}}$ is expected to be a
Kosterlitz-Thouless
transition point.
The nature of the operator $\tilde{O}(x)$ and, accordingly, the type of
ordering in a massive phase which occurs at $\Delta < - \frac{1}{\sqrt{2}}$ can
be elucidated
by means of the effective sine-Gordon model written in terms of the dual field
$\tilde{\phi}$
\begin{equation}
H_{eff} = \int dx [ \frac{u^{'}}{2} \left( \tilde {P}^{2}(x) +
(\partial_{x} \tilde {\phi} (x) )^{2} \right)
- \frac{m}{\alpha} \cos 2 \tilde{\beta} \tilde{\phi}(x) ],
\label {eq: SG}
\end{equation}
together with the continuum representation of spin operators in terms of scalar
fields
$\phi$ and
$\tilde{\phi}$ \cite {LP,Shan}:
\begin{equation}
S^{z}_{n} \rightarrow \frac{\beta}{2\pi} \partial_{x} \phi (x) +
\lambda (-1)^{n} \sin \beta \phi (x),
\label {eq: S1}
\end{equation}
\begin{equation}
S^{x}_{n} \rightarrow \mu (-1)^{n} \cos \frac{1}{2} \tilde{\beta}
\tilde{\phi}(x), ~~~~
S^{y}_{n} \rightarrow \mu (-1)^{n} \sin \frac{1}{2} \tilde{\beta}
\tilde{\phi}(x)
\label {eq: S2}
\end{equation}
In the above equations, $m$ is a positive parameter proportional to $\delta^{2}
a / u$, and
$\lambda$ and $\mu$ are numerical constants. As follows from (\ref {eq: SG}),
the onset of a strong-coupling
regime in the model (\ref {eq: sdmod}) results in ordering of the dual field,
$\tilde{\phi}$.

Let us look at the symmetry properties of the model (9) under translations and
spin
rotations about the $\hat{z}$-axis \cite {Affl}. The spin operators in (10) and
(11)
are invariant under
$
\phi \rightarrow \phi + 2\pi/\beta,$
$\tilde{\phi} \rightarrow \tilde{\phi} + 4\pi/\tilde{\beta}.$
This is the general symmetry of any spin model. Under a translation by one
lattice spacing
$(T_a),$
$
\phi \rightarrow \phi + \pi/\beta,$
$\tilde{\phi} \rightarrow \tilde{\phi} + 2\pi/\tilde{\beta}.$
Under a U(1) rotation about the $\hat{z}$-axis by an angle $\alpha$
$(U_{\alpha}),$
$\phi$ does not change, while
$
\tilde{\phi} \rightarrow \tilde{\phi} + 2\alpha/\tilde{\beta}.$
Using these transformations of the fields $\phi$ and $\tilde{\phi}$, one can
check that
the perturbation term in the original continuum model (\ref {eq: sdmod})
changes its sign
under $T_a$ and $U_{\pi/2}$, thus reflecting breakdown of translational
symmetry
(bond alternation) and the $\pi/2$ spin rotation symmetry (XY-anisotropy).  We
then notice
that, in the effective model (\ref {eq: SG}), each of these two symmetries is
recovered;
however
the operator $\cos 2 \tilde{\beta} \tilde{\phi}$
changes its sign under $U_{\pi/4}$. We therefore conclude that
the effective perturbation, generated in the second order in
$\delta$, represents a {\it translationally invariant, two-axis} ($Z_4$)
anisotropy. Its
lattice version is given by a four-spin term
$$\cos 2 \tilde{\beta} \tilde{\phi} \leftrightarrow
(S^{x}_{n} S^{x}_{n+1} - S^{y}_{n} S^{y}_{n+1})
(S^{x}_{n+2} S^{x}_{n+3} - S^{y}_{n+2} S^{y}_{n+3})$$
$$
- (S^{x}_{n} S^{y}_{n+1} + S^{y}_{n} S^{x}_{n+1})
(S^{x}_{n+2} S^{y}_{n+3} + S^{y}_{n+2} S^{x}_{n+3}).
$$

In the strong-coupling phase $(-1 < \Delta < - 1/\sqrt{2})$ the $Z_4$-symmetry
is broken.
The field $\tilde{\phi}$
acquires a nonzero vacuum expectation value, determined by one of the four
degenerate minima
$\tilde{\phi} = 0, ~\pi/\tilde{\beta}, ~2\pi/\tilde{\beta},
{}~3\pi/\tilde{\beta}$ of the
"potential" $- m \cos 2 \tilde{\beta} \tilde{\phi}$
(within the main periodicity interval $4\pi/\tilde{\beta}$). As follows from
(\ref {eq: S2}),
this leads to  a Neel ordering of the spins, either along the $x$-axis,
\begin{equation}
(-1)^{n}<S^{x}_{n}> \sim \pm \mu,~~(-1)^{n}<S^{y}_{n}> = 0 \label {eq: ord1}
\end{equation}
at $\tilde{\phi} = 0, 2\pi/\tilde{\beta}$, or along the $y$-axis,
\begin{equation}
(-1)^{n}<S^{x}_{n}> = 0, ~~(-1)^{n}<S^{y}_{n}> = \sim \pm \mu \label {eq: ord2}
\end{equation}
at $\tilde{\phi} = \pi/\tilde{\beta}, 3\pi/\tilde{\beta}$.

As we already mentioned, the results obtained for a weak bond-alternating
anisotropy
can be applied to the opposite limiting case $|\delta| \gg J$, using the
$\delta \leftrightarrow J$ symmetry (\ref {eq: sym3}), (\ref {eq: twstr}).
In the region
$-1 < \Delta' < 1/\sqrt{2}$, where $\Delta' = J_z/|\delta|$, we find two
gapless SN phases
(for each sign of $\delta$), with low-temperature properties of a Luttinger
liquid.
Applying nonlocal transformations (\ref {eq: twstr}) to Eqs.(\ref {eq: ord1})
and
(\ref {eq: ord2}), we find that at $\Delta' =1/\sqrt{2}$ a Kosterlitz-Thouless
transition
takes place to an ordered phase, characterized by the
$\uparrow \uparrow \downarrow \downarrow$ periodic
long-range alignment of the spins, either along the $x$- or $y$-direction in
spin space.

\section {Conclusions}

We have shown in this paper that, in the $S=1/2$ quantum spin chains,
the gapless, Luttinger spin-liquid state with
a power-law decay of the correlation functions is not exhausted by systems
possessing
translational and spin rotation symmetries. Specifically,
this type of low-energy behavior also characterizes  the XXZ spin chain with
the
staggered XY anisotropy, exhibiting XY-like and spin-nematic phases with a
gapless excitation spectrum. The model possesses a number of interesting
properties.
We have shown that massless Majorana fermions with
different velocities, being elementary excitations in the noninteracting case
of
the XY bond-alternating chain, fail to be observable (confinement) in the
Luttinger-liquid
regime, when interaction caused
by a finite $zz-$anisotropy $\Delta$ is switched on. Another interesting
features
of the model are order from disorder and Kosterlitz-Thouless transitions
to massive ordered phases, driven by the $\delta-$anisotropy.

Although
our description was restricted by limiting cases $|\delta| \ll J$ and
$|\delta| \gg J$, it is clear that the gapless XY-like and SN phases occupy
large
domains in the parameter space of the model.
However, to understand the phase diagram of
the system in more detail and, in particular, describe  a transition from the
XY-like
phase to the SN phases on changing $\delta$ with $\Delta$ kept finite,
 the region $|\delta| \sim J$ should be considered.
This region is not accesible by bosonization method.
As follows from the two-chain representation (\ref {eq: 2sdchain}) of the
original model
(\ref{eq:basic}),
in this region one has to consider an interacting system of "light" and
"heavy" Majorana fermions with strongly different velocities $(v_+ \gg v_-)$.
This problem resembles the Kondo-lattice one and deserves a special analysis.
We hope to return to this and related questions in the future.

\acknowledgments
We thank P.Chandra,  A.Finkelstein, S.Girvin,  H.Johannesson,
 S.\"{O}stlund and especially A.Tsvelik
for interesting discussions and helpful comments. A.N. would like to
gratefully acknowledge
the financial support from Chalmers University of Technology and Nordita.


\begin{references}
\bibitem[*]{email} e-mail: ners@fy.chalmers.se
\bibitem {Hida} K.Hida, Phys.Rev. {\bf B 45} (1992) 2207;  {\bf B 46} (1992)
8268.
\bibitem {Kom-Tas} M.Kohmoto and H.Tasaki,  Phys.Rev. {\bf 46} (1992) 3486.
\bibitem {Yam} M.Yamanaka, Y.Hatsugai and M.Kohmoto, preprint.
\bibitem {Hald-gap} F.D.M.Haldane, Phys.Lett. {\bf 93 A} (1983) 464;
Phys.Rev.Lett.{\bf 50}
(1983) 1153.
\bibitem {XYZ} J.D.Johnson, S.Krinsky amd B.M.McCoy,  Phys.Rev.{\bf A 8} (1973)
2526.
\bibitem {A-G} A.F.Andreev and I.A.Grishchuk, Sov.Phys.- JETP {\bf 60} (1984)
267.
\bibitem {Vil} J.Villain, J.Physique {\bf 38} (1977) 26; J.Villain, R.Bidaux,
I.P.Carton and
R.Conte, J.Physique {\bf 41} (1980) 1263.
\bibitem {KLN} F.V.Kusmartsev, A.Luther and A.A.Nersesyan, JETP Lett. {\bf 55}
(1992) 692.
\bibitem {conf}  P.W.Anderson,  Phys.Rev.Lett. {\bf 67} (1991) 3844.
\bibitem {KNK} M.Kohmoto, M.den Nijs and L.P.Kadanoff, Phys.Rev.{\bf B 24}
(1981) 5229.
\bibitem {ID} See, for instance, in "{\it Statistical Field Theory}"
by C.Itzykson and J.-M.Drouffe, v.1, Chapter 2, Cambridge University Press,
1989.
\bibitem {XXZ} C.N.Yang and C.P.Yang, Phys.Rev.{\bf 150} (1966) 321.
\bibitem {Bax} R.J.Baxter, {\it Exactly Solved Models in Statistical
Mechanics},
Academic Press,
1982.
\bibitem {SCZ} S.-C.Zhang,  Phys.Rev.Lett.{\bf 65} (1990) 120;
S.\"{O}stlund and E.Mele, Phys.Rev.{\bf B 44} (1991) 12413.
\bibitem{foot}
For a given $q$, local Fock space of states is 4-dimensional, with basis
vectors $|n_{q}, n_{\pi-q}>$, $n_{q} = 0,1$ being the occupation number.
The SU(2) algebra generated by $J^{\alpha}_{q}$ in (\ref {eq: gen}),
$[J^{\alpha}_{q}, J^{\beta}_{q^{'}}] = i \delta_{q, q^{'}}
\epsilon^{\alpha \beta \gamma} J^{\gamma}_{q}$, is defined in a 2-dimensional
subspace with basis vectors $|00>$ and $|11>$. A particle-hole transformation
$a_{\pi-q} \leftrightarrow a^{+}_{\pi-q}$ maps this subspace onto the
orthogonal one, with basis vectors $|10>$ and $|01>$. The transformed operators
$J^{\alpha}_{q}$ generate one more SU(2) algebra. So, the total local symmetry
group of $H_{0}$ is $SU(2) \otimes SU(2)$.
\bibitem {LSM} E.Lieb, T.Schultz and D.Mattis, Ann.Phys. {\bf 16} (1961) 407.
\bibitem {CCL1} P.Chandra, P.Coleman and A.I.Larkin, J.Phys.: Condens.Matter,
{\bf 2} (1990) 7933.
\bibitem {CCL2} P.Chandra, P.Coleman and A.I.Larkin, Phys.Rev.Lett.{\bf 64}
(1990) 88.
\bibitem {LP} A.Luther and I.Peschel, Phys.Rev.{\bf B 12} (1975) 3908.
\bibitem {Hald-LL} F.D.M.Haldane, Phys.Rev.Lett.{\bf 45} (1980) 1358; {\bf 47}
(1981) 1840.
\bibitem {Shan} R.Shankar, Int.J.Mod.Phys. {\bf B 4} (1990) 2371.
\bibitem {Umkl} J.L.Black and V.J.Emery, Phys.Rev.{\bf B 23} (1981) 429;
M.den Nijs Phys.Rev.{\bf B 23} (1981) 6111.
\bibitem {Yak} V.M.Yakovenko, JETP Lett. {\bf 56} (1992) 5101.
\bibitem {NLK} A.A.Nersesyan,  A.Luther and  F.V.Kusmartsev, Phys.Lett. {\bf A
176}
(1993) 363.
\bibitem {spin}
{A.N. is indebted to Alexei Tsvelik for clarifying discussions on this point.}
\bibitem {Affl} S.Eggert and I.Affleck,  Phys.Rev. {\bf B 46} (1992) 10866.

\end{references}
\end{document}